\documentclass[fleqn,10pt]{wlscirep}
\usepackage{ulem}
\usepackage{lineno}

\title{Homeostatic plasticity and emergence of functional networks in a whole-brain model at criticality}

\author[1,2,3,*]{Rodrigo P. Rocha}
\author[2,3]{Loren Ko\c{c}illari}
\author[2,3]{Samir Suweis}
\author[3,4,5]{Maurizio Corbetta}
\author[2,3]{Amos Maritan}
 
\affil[1]{Department of Physics, School of Philosophy, Sciences and Letters of Ribeir\~{a}o Preto, University of S\~{a}o Paulo, 
Ribeir\~{a}o Preto, SP, Brazil.}
\affil[2]{Dipartimento di Fisica e Astronomia, Universit\`a di Padova and INFN, via Marzolo 8, I-35131 Padova, Italy.}
\affil[3]{Padova Neuroscience Center, Universit\`a di Padova, Padova, Italy.}
\affil[4]{Dipartimento di Neuroscienze, Universit\`a di Padova, Padova, Italy.}
\affil[5]{Departments of Neurology, Radiology, Neuroscience, and Bioengineering, Washington University, School of Medicine, St. Louis, USA.}
 
\affil[*]{Correspondence to rodrigo.rocha@ufsc.br}


\begin{abstract}
Understanding the relationship between large-scale structural and functional brain networks remains a crucial issue in modern neuroscience. Recently, 
there has been growing interest in investigating the role of homeostatic plasticity mechanisms, across different spatiotemporal scales, in regulating network activity 
and brain functioning against a wide range of environmental conditions and brain states (e.g., during learning, development, ageing, neurological diseases). In the 
present study, we investigate how the inclusion of homeostatic plasticity in a stochastic whole-brain model, implemented as a normalization of the incoming node's 
excitatory input, affects the macroscopic activity during rest and the formation of functional networks. Importantly, we address the structure-function relationship both at 
the group and individual-based levels. In this work, we show that normalization of the node's excitatory input improves the correspondence between simulated neural patterns of the 
model and various brain functional data. Indeed, we find that the best match is achieved when the model control parameter is in its critical value and that normalization 
minimizes both the variability of the critical points and neuronal activity patterns among subjects. Therefore, our results suggest that the inclusion of homeostatic principles 
lead to more realistic brain activity consistent with the hallmarks of criticality. Our theoretical framework open new perspectives in personalized brain modeling with 
potential applications to investigate the deviation from criticality due to structural lesions (e.g. stroke) or brain disorders.
\end{abstract}

\begin{document}

\flushbottom
\maketitle
\thispagestyle{empty}

\section*{Introduction}

The human brain constitutes an impressively complex system characterized by many spatiotemporal scales. At the large-scale, white matter pathways derived from diffusion tensor 
or diffusion spectrum imaging (DTI/DSI) define the so-called human connectome \cite{Kotter2005}, i.e., a structural network of hard-wired interconnections among mesoscopic 
brain regions. On the other hand, large-scale brain activity can be accessed, among other techniques, through functional magnetic resonance imaging \cite{Pol2010} (fMRI) which 
is a four-dimensional and non-invasive imaging technique that measures changes in the blood oxygen level dependent (BOLD) over time. BOLD time-series behave as spontaneous 
low-frequency ($<0.1$ Hz) fluctuations that have been shown to be highly correlated across different brain areas, at rest or during a given cognitive task \cite{Sporns2009,Friston2013,Raichle2005,Beckmann2006}. 
Structural connections are the main input of whole brain models \cite{Cabral2014,Deco2017} that have been developed for understanding how patterns of correlated activity 
among brain regions, also called functional connectivity (FC), emerge. Understanding the relationship between structural and functional connectivity remains a crucial issue in modern neuroscience, 
and many studies focus on developing methods to increase the similarity between simulated and empirical functional activity, using the connectome as input 
\cite{HagmannPNAS2009,Raj2014,Jirsa2014,Messe2015,Jirsa2016}.

From a theoretical point of view, statistical physics has decisively contributed to highlight the potential advantage that a brain may have in a critical state and also provide 
a quantitative  description of brain activities though minimalist mesoscopic models \cite{Chialvo2010,Hai2013}. Systems consisting of many microscopic components (e.g. neurons) 
may exhibit rather diverse types of macroscopic collective behavior with different levels of internal organization (e.g. brain activity). Moreover, slight changes in 
external stimuli (e.g. auditory, visual , etc.) or in the strength of interactions themselves may induce dramatic structural rearrangements, i.e. phase transitions. It is thus 
tempting to hypothesize that biological states might be manifestations of similar collective phases and that shifts between them could correspond to phase transitions. 

The emerging hypothesis is that living systems, or parts of them, like the brain, are spontaneously driven close to a critical phase transition 
\cite{obs1,HidalgoPNAS2014,Munoz2017}, thus conferring upon them the emergent features of critical systems like the lack of spatial and temporal scales and the high responsiveness 
to external perturbations. These characteristics would translate into the ability of the brain, through a large spatial and temporal scale activity, to promptly react to 
external stimuli by generating a coordinated global behavior \cite{Schneidman2006}, to maximize information transmission \cite{Beggs2003,Larralde2017}, sensitivity to sensory 
stimuli \cite{Kinouchi2006} and storage of information \cite{Beggs2005}.

These ideas have been particularly investigated in the last fifteen years in neuroscience and the hypothesis that the brain is poised near a critical state (in statistical 
mechanics sensu) is gaining consensus in the neuroscience community \cite{Chialvo2010,SpecialIssue,Hesse2014,Munoz2017,Cocchi2017}. In brain systems, the concept of criticality is 
mainly supported by the following two experimental findings: (i) the discovery of scale-free neural avalanches \cite{Beggs2003}, as described by power-law distributions for 
the size and duration of the spontaneous bursts of activity in the cortex;  (ii) the presence of long-range temporal correlations in the amplitude fluctuations of neural 
oscillations \cite{Hansen2001,Hardstone2012}. Further studies reported the universality of the power-law exponents originally found in \cite{Beggs2003} among different 
species, for instance, rat \cite{Gireesh2008}; non-human primate \cite{Petermann2009,Yu2011} and humans using diverse techniques, such as MEG \cite{Poil2012,Palva2013,Shriki2013}; 
EEG \cite{Meisel2013} and fMRI \cite{Tagliazzuchi2012,Hai2013}.

Also from a theoretical point of view, many whole brain models maximally describe real-neuronal activities when they are poised at a critical point 
\cite{Beggs2003,ChialvoPRE,Hai2013,Deco2012,JirsaDeco2014,Tagliazzuchi2012}. Recently, a whole-brain mesoscopic model (which we call here HTC model), proposed by Haimovici {\it et. al.} 
\cite{Hai2013}, which is a variant of the Greenberg-Hastings cellular automata \cite{Greenberg78}, predicts a phase transition between the sub-critical regime with low 
activity, and the super-critical regime of high activations. When poised at the critical point, the HTC model \cite{Hai2013} is able to capture, at the group level, 
the emergence of experimental spatiotemporal patterns, the temporal correlation between regions (functional connectivity, FC), the organization of brain wide patterns in so 
called resting state networks (RSNs), the scaling law of the correlation length, among others. Typically these studies have been used to investigate healthy brain activity at the group level (using a single averaged functional and 
structural matrix from a cohort of healthy subjects) while little attention has been given to unhealthy brains \cite{SpecialIssue,Hesse2014,Cocchi2017}. In particular, personalized brain modelling 
(which uses single individual DTI and fMRI as model input) has been largely unexplored for both healthy and unhealthy brains.

Recent experimental findings suggest that brain diseases (e.g., injuries, disorders) could promote a departure from the critical regime, as reported in studies of anesthesia 
\cite{Scott2014}, slow wave sleep \cite{Priesemann2013} and epilepsy \cite{Meisel2012}, where fundamentally deviations from healthy conditions promote a loss in long-range 
correlations and power-law distributions for the (spatiotemporal) neural avalanches. From a theoretical point of view, a recent work by Haimovici {\it et. al} \cite{Hai2016} 
has quantified the way synthetic lesions may impact the large-scale dynamical signatures of the (HTC) critical dynamics at a group level. Synthetic lesions are able to push 
the system out of the critical state towards a sub-critical state, which is characterized by decreased levels of neural fluctuations. Sub-critical dynamics 
also lead to alterations in the functional parameters, for instance, the mean and variance of the FC matrix are decreased due to synthetic lesions. These results are in 
agreement with previous experimental and theoretical studies \cite{Scott2014,Priesemann2013,Meisel2012,Sporns2008,Alstott2009,Kringelbach2014,Hellyer2015,Roy2016,Raj2016,Marinazzo2016,Adhikari2017,Saenger2017}. 
However, such an approach only holds for averaged groups, because the critical point in the HTC model is subject dependent resulting in a very high inter-subject variability, thus 
inhibiting the possibility to distinguish markers of neural activity and functional patterns between healthy and injured brains. 

The general concept of systems tuning themselves to critical states is known as self-organized criticality (SOC) \cite{Bak1988}. The observed stability of 
the neural activity against large perturbations, such as abrupt changes in environmental conditions and brain states (e.g., during learning, development, ageing, 
neurological diseases) is widely believed to be maintained by an array of Hebbian-like and homeostatic plasticity mechanisms that regulate neuronal and circuit 
excitability \cite{Desai2010,Turrigiano2011}. It has been suggested that these mechanisms play a crucial role in the brain's criticality \cite{Hesse2014}. Indeed, the 
self-organization of biologically relevant neural models to criticality has been investigated in a number of studies with varying degrees of sophistication. Diverse forms of 
plausible synaptic plasticity mechanisms have been analyzed at the microscopic level, such as, activity-dependent rewiring \cite{Rohl2003,Tetzlaff2010}, Hebbian plasticity \cite{Arcangelis2006}, short-term synaptic 
plasticity \cite{Geisel2007,Geisel2009,Millman2010}, spike timing dependent plasticity (STDP) \cite{Rubinov2011,Gross2009}, and homeostatic plasticity \cite{Droste2013}. 
Nevertheless, the biological mechanisms underlying the self-organization at a macroscopic scale remains unclear. Recent theoretical work suggest that homeostatic plasticity 
mechanisms may play a role in facilitating criticality, hence the emergence of functional brain networks at the 
macroscale \cite{Corbetta2014,Leech2015,Leech2017}.

Here we model the putative role of homeostatic plasticity mechanisms in regulating brain activity, criticality, and brain networks. Specifically, we introduce a variant of 
the stochastic HTC model in which we introduce a normalization of the structural connectivity matrix that effectively equalizes the excitatory input, i.e. it maintains the 
original topology while rescaling the weights of existing connections. Therefore, the implemented normalization acts as a homeostatic plasticity principle balancing network 
excitability. We show that the inclusion of homeostatic mechanisms leads to more realistic brain activity consistent with the hallmarks of criticality. Indeed, normalization of 
the node's excitatory input improves the correspondence between simulated neural patterns of the model and various brain functional data, such as the functional connectivity (FC), 
resting state networks (RSNs) and the power-law distribution for the sizes of active clusters in the cortex. An important result of the proposed framework is that we are able to 
reduce the inter-subject variability within the class of healthy brains. In particular, we show that network normalization collapses the model state variables, i.e. neural 
activity patterns, of healthy subjects into universal curves, opening up a potential application on personalized brain modeling.

\section*{Theoretical Framework}
The HTC model \cite{Hai2013} consists of a discrete three-state cellular automaton in a network of $N$ nodes (i.e. cortical brain regions) linked with symmetric and 
weighted connections obtained from DTI/DSI scans of the white matter fiber tracts \cite{DTI_REVIEW} and described by a matrix $W$. The diagonal elements of $W$ 
(i.e., self-connections) are all set to zero. At any given time step, each node can be in one of the three possible states: active ($A$), inactive ($I$), and refractory 
($R$). The state variable of a given node $i$, $s_i(t)$, is set to $1$ if the node is active and $0$ otherwise. The temporal activity of the $i$-th node is governed by the 
following transition probabilities between pair of states: (i) $I \to A$ either with a fixed probability $r_1$ or with probability $1$ if the sum of the connections weights 
of the active neighbors $j$,  $\sum_j W_{ij}$, is greater than a given threshold $T$, i.e., $\sum_j W_{ij}s_j >T$, otherwise $I \to I$, (ii) $A \to R$ with probability $1$, 
and (iii) $R \to I$ with a fixed probability $r_2$. The state of each node is overwritten only after the whole network is updated. The two parameters $r_1$ and $r_2$ controls 
the time scale of self-activation and recovery of the excited state, while $T$ sets the rate of induced activity due to active nearest neighbors \cite{Hai2013}. 

A mean field version of the dynamics is easily obtained in terms of the probability of node $i$ to be active, $p_i^t$, or quiescent, $q_i^t$, or refractory, 
$r_i^t=1-p_i^t-q_i^t$ (not to be confused with $r_1$ and $r_2$ which are the model parameters):
\begin{eqnarray}
p_i^{t+1}&=&q_i^t\Big[ r_1 + (1-r_1)\Theta(\sum_{j=1}^N W_{ij}p_j^t-T) \Big], \label{p}\\
q_i^{t+1}&=&q_i^t+r_2(1-p_i^t-q_i^t)-p_i^{t+1}, \label{q}
\end{eqnarray}
where $\Theta$ is the Heaviside unit step function. In Eq. \eqref{p} we assumed the neighbors of node $i$ being excited as independent events. As discussed earlier 
\cite{PRL2011}, this approximation yields good results even when the network has a non-negligible amount of short loops, which is the case of DTI/DSI networks considered 
in this study. Analytical solutions for $p_i^t$ and $q_i^t$ are difficult to be obtained. However, under suitable considerations one can obtain an approximate solution for 
the critical point $T_c$ (see Methods section), that explains its high variability within subjects. In addition, as we show in the Methods section, Eqs. (\ref{p})-(\ref{q}) 
correctly predict a collapse of $T_c$ across subjects when the normalized version of the input matrix $W$ is used to simulate the dynamics.

The complex behavior of the functional activities of the human brain is thought to emerge by the underlying architecture of the anatomical brain connections, as given 
by the binary adjacency matrix of the human connectome. In order to consider homeostatic principles regulating network excitability, we introduce a 
normalization of the structural connectivity matrix. Indeed, previous results \cite{Odor2016} have shown that simulated mesoscopic neuronal network dynamics is dominated 
by the central nodes, i.e. hubs with high in-degree strength $W_i=\sum_j W_{ij}$. In order to regulate network excitability, following \cite{Odor2016,Lesne2012}, 
we here propose a variant of the HTC model, by normalizing locally each entry of the structural matrix according to the following normalization rule:
\begin{equation}
\widetilde{W}_{ij}=W_{ij}/\sum_j W_{ij}.
\label{equalization}
\end{equation}
The motivation behind Eq. \eqref{equalization} is as follows. In the HTC model, a node activation happens when the incoming input excitation from its nearest 
active neighbors exceeds a fixed threshold $T$, i.e, $\sum_j W_{ij}s_j > T$. In this way, one may interpret $T$ as a threshold parameter that regulates the propagation of 
incoming excitatory activity (similar to an action potential in spiking neuron models). In biological terms, normalization could be viewed as a homeostatic plasticity 
principle aiming to regulate excitation and inhibition through the balancing of the structural weight connections. Indeed, it fixes the in-degree of all nodes to $1$, ensuring that each 
node has at the mesoscopic level a similar contribution on regulating the simulated brain activity \cite{obs3}.

In the numerical simulation we have discretized time in steps $dt$. We set the total simulation time-steps $t_s$, so to recover the length of typical (fMRI) BOLD 
experimental time-series (about 5-20 minutes). In order to characterize simulated brain  activity we have analyzed some standard quantities (see Methods section): the mean 
network activity ($\langle A \rangle$), the standard deviation of $A(t)$ ($\sigma (A))$ and the sizes of the averaged clusters, the largest $\langle S_1\rangle$ and the 
second largest $\langle S_2 \rangle$. The clusters of activity were defined as the size of the connected components of the graph defined by the sets of 
nodes that are structurally connected to each other and simultaneously active.

The simulated dynamics displays a phase transition as $T$ varies while keeping $r_1$ and $r_2$  fixed. For small values of the activation threshold $T$, the activity is 
over-responsive and the signal from an active node will spread all over its first neighbors. We refer to this phase as a super-critical phase, which is characterized by 
sustained spontaneous activity with fast and temporally  uncorrelated fluctuations. On the other hand, high values of $T$ leads to a sub-critical phase, which is 
characterized by regular, short propagating and not self-sustained  brain activity. In this phase, only those nodes with the strongest connections will determine the 
signal flow in the network. In between of these two phases a phase transition occurs at $T=T_c$ where brain activities have oscillatory behaviors, and long-range temporal 
correlations in their envelope \cite{Hai2013,Hai2016}. As shown in \cite{Hai2013} the size of the second largest cluster is a suitable quantity to characterize the phase 
transition and it happens at the corresponding value of $T$ where $\langle S_2\rangle$ is maximal (see also \cite{Stauffer1982}). In addition to the second cluster size, the peak in the standard deviation, 
$\sigma(A)$, may also be used to infer the critical transition \cite{Hai2016}. 

To address the effects of homeostatic principles in whole-brains, we thus performed our analysis using as input both structural matrices, $W$ and in its normalized counterpart 
$\widetilde{W}$. Herein we show that our approach is able to capture, at the critical point, the emergence of functional connectivity at rest, resting state networks (RSNs), 
among others. In particular, we find that the HTC model leads to more realistic predictions when the normalization is considered and, in this case, it can also be 
successfully applied to individual personalized brain analysis.

\section*{Results}

\subsection*{Group brain modelling: using average connectome as model input}

\textbf{Hagmann et al. dataset.} We first compare the output of the presented whole-brain model on a low-resolution structural network with $N=66$ cortical regions obtained 
as an average connectome of 5 individuals \cite{Hag2008}. The advantage of working with this dataset is that we have both average structural and functional networks, and we 
have a reference template for the resting state networks (see Methods). On the other hand, the DTI/fMRI matrix for each single individuals is missing, and therefore, in this case, 
we cannot perform an individual brain modelling.

We fixed the model parameters to the following values: $r_1=2/N$, $r_2=r_1^{1/5}$, $t_s=6,000$ time-steps with time discretized in $dt=0.1$ seconds. We arbitrarily chose 
this parameterization in order to keep the ratio $r_1/r_2$ similar to that used originally in \cite{Hai2013}. We then computed $\langle A\rangle$, 
$\sigma(A)$, $\langle S_1\rangle$ and $\langle S_2\rangle$, as a function of the threshold $T$ in the interval $[0,0.3]$. We used black (red) color to represent the input matrix $W$ 
($\widetilde{W}$). 

\begin{figure}
\centering
\includegraphics[width=1 \linewidth]{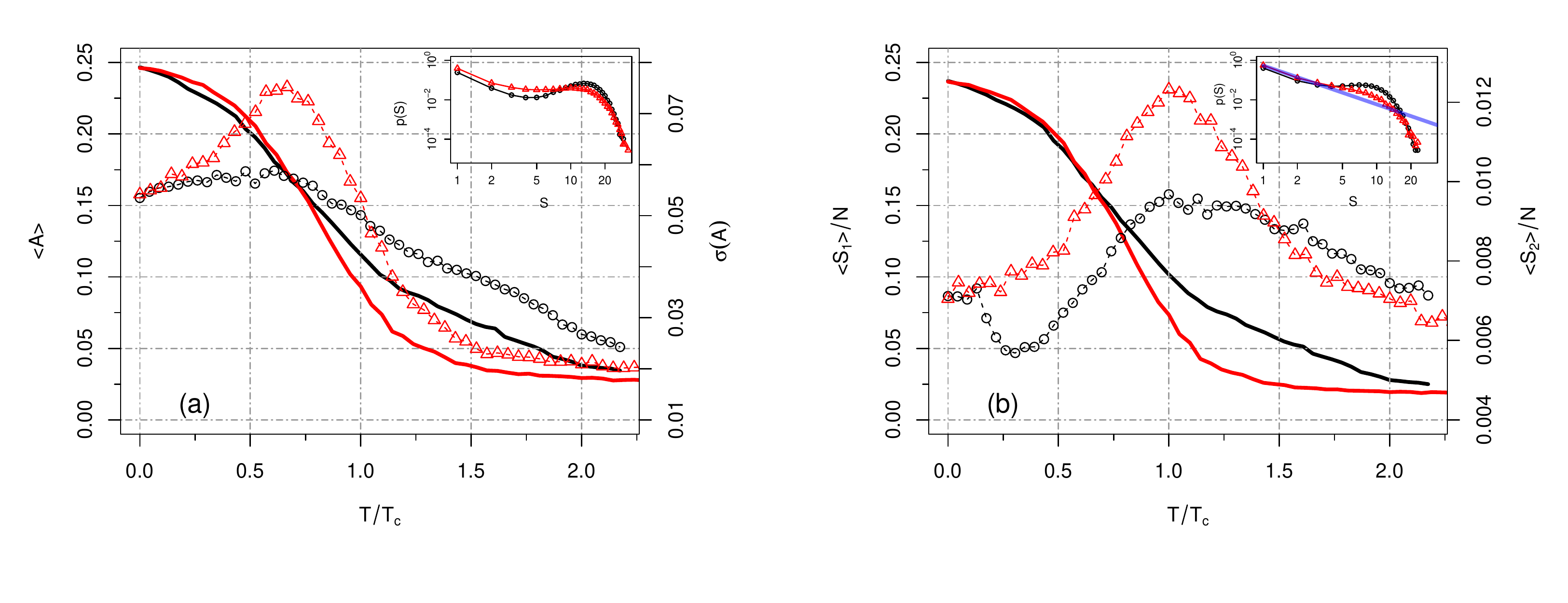}
\caption{Results of our framework applied to the Hagmann {\it et. al} dataset \cite{Hag2008}. (a) Mean activity (solid lines) and its standard deviation (dots) 
as a function of $T/T_c$. The critical points $T_c$, for all cases, are set as the value corresponding to the maximum of $\langle S_2\rangle$ because of the power-law 
distribution of cluster sizes found at this point. In black we depict the results for the non-normalized matrix $W$, while in red we depict the normalized matrix 
$\widetilde{W}$. Inset: Distribution of cluster sizes (in log-log scale) for the corresponding peaks of $\sigma(A)$. Scale invariance is not as visible as at the peak of 
$\langle S_2\rangle$. (b) Largest (solid lines) and second largest cluster (dots). The major effect of equalization of the network sensitivity is to enhance the strength of 
spatiotemporal variability, as seen in the peaks of $\langle S_2 \rangle$ and $\sigma(A)$. Inset: Distribution of cluster sizes (in log-log scale) for the corresponding 
peaks of $\langle S_2\rangle$. The blue solid is the fit of Eq. \eqref{CDF} used to estimate $\alpha=1.97\pm 0.03$ for the normalized network.}
\label{FIG1}
\end{figure}

In Fig. \ref{FIG1} we show $\langle A\rangle$ (solid lines) and $\sigma(A)$ (dots) as a function of the rescaled threshold $T/T_c$, where $T_c$ corresponds to the maximum of 
$\langle S_2\rangle$. Interestingly, the major global effect of normalization (for a fixed $T$) is not to increase the mean network activity $\langle A \rangle$, which in 
turn remains almost unchanged for small values of $T$, but to distribute the activity more evenly across the network. Accordingly, we observe an overall increase in the 
strength of spatiotemporal neural variability as revealed by the peaks of $\langle S_2\rangle$ and $\sigma(A)$. In fact, both peaks become more pronounced after the network 
normalization. Finite size systems show a smooth behavior in correspondence of a phase transition in the infinite size system. Thus the divergence of a susceptibility at a 
critical point of an infinite system becomes a smooth peak for the corresponding finite system \cite{obs2}. These peaks of maximum variability happen for different values 
of $T$, but this effect is due to finite size effects (small $N$) affecting the position of the critical point in the system. The critical thresholds for the two quantities 
are expected to converge to the same $T_c$ as $N\rightarrow \infty$ and this expectation is confirmed when using $N$ large enough (see the next section).

Another signature of criticality in the system is the cluster sizes distribution, $p(S)$. As shown in \cite{Tagliazzuchi2012,Hai2013} the brain forms activity clusters whose 
sizes follows a truncated power-law distribution, i.e. $p(S)\sim S^{-\alpha}\exp(-S\gamma)$, with the exponential cut-off due to finite size ($\gamma\propto1/N$) and a power-law 
slope of $\alpha=3/2$ consistent with the hallmark exponent of neuronal avalanches \cite{Beggs2003}. Notice that $\gamma$ should diverge at $T_c$ when $N\rightarrow \infty$. 
In order to identify the critical point of the system for both $W$ and $\widetilde{W}$ in the case of small $N$, we compute the distribution of cluster sizes for some values 
of $T$, including those corresponding to the peaks of $\sigma(A)$ and $\langle S_2\rangle$ (see insets of Fig. \ref{FIG1}). 
Indeed, at the peak of the second largest cluster, we find a truncated power law-distribution, with an exponent $\alpha=1.97\pm 0.03$ and a cutoff which depends on the mean 
activity level \cite{Tagliazzuchi2012,Hai2013} (see the Methods section for the fitting procedure). Furthermore, the normalized dynamics with $\widetilde{W}$, shows a 
better power-law distribution, extending up to cluster sizes $S\approx 12$, than the non-normalized dynamics, $W$, where it extends up to $S\approx 5$ . On the other hand, 
scaling (if any) is less visible for $T$ corresponding to the peak of $\sigma(A)$ (see inset of Fig \ref{FIG1} (a)). Therefore from now on, we will define the critical point $T_c$ at 
the maximum of the second largest cluster. We finally note that the average size of the largest cluster (Fig. \ref{FIG1} (b) - solid lines) is almost indistinguishable from the 
average activity (Fig. \ref{FIG1} (a) - solid lines) in the full range of thresholds considered. This result shows that most of the time the active regions form patterns 
organized in a single giant component. Indeed, the largest cluster is almost two orders of magnitude larger than the second largest cluster. 

\begin{figure*}
\centering
\includegraphics[width=\linewidth]{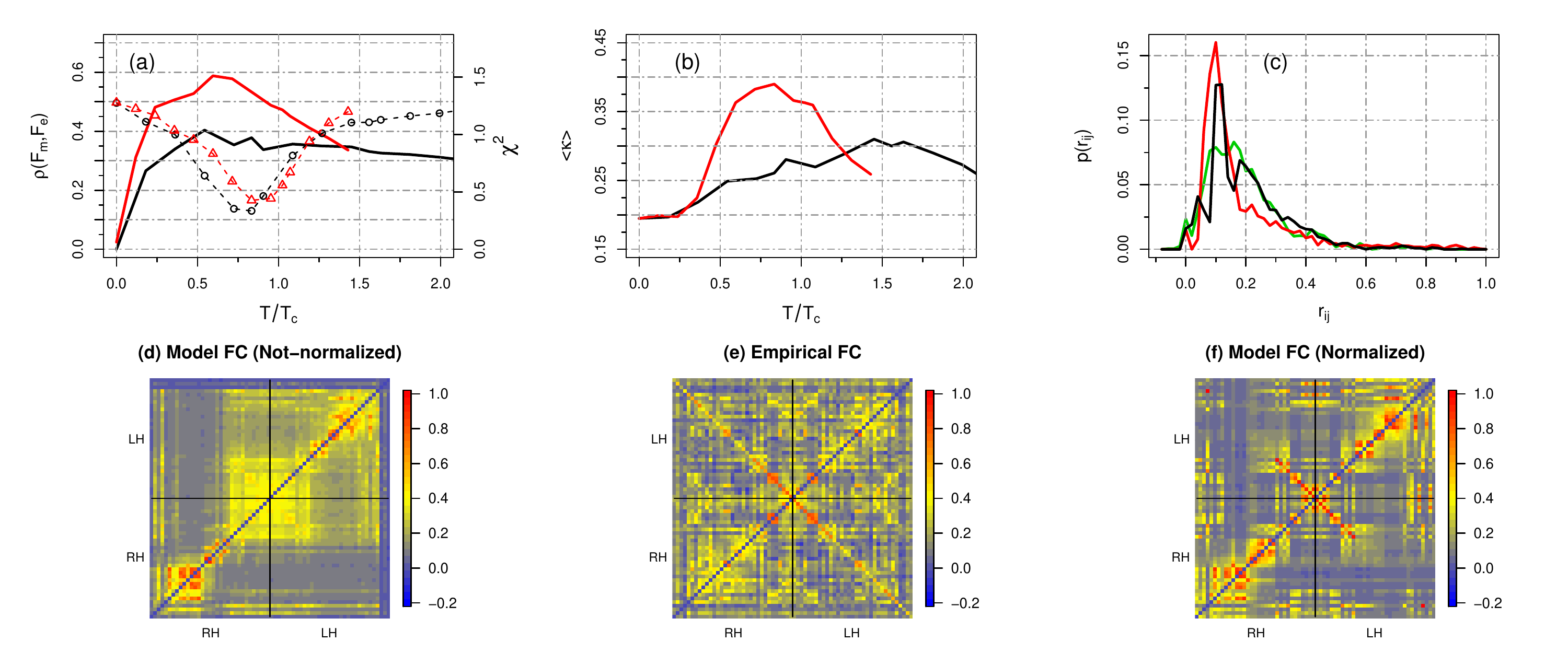}
\caption{(a) Pearson correlation (solid lines) between simulated and empirical functional connectivity matrices, $\rho(F_m,F_e)$, as a function of $T/T_c$ (black color for 
$W$ and red color for $\widetilde{W}$). Chi-squared distance (dots) between the corresponding (normalized) probability distribution functions. The normalization of the excitatory 
input ($\widetilde{W}$) enhances the match between empirical and simulated data by a factor of about $1.5$. The best match ($\rho\approx 0.6$) occurs at $T$ corresponding to the 
peak of $\sigma(A)$, while the smallest distance, $\chi^2 \approx 0.4$, occurs at $T$ corresponding to the peak of $\langle S_2\rangle$. (b) Overall match between empirical 
resting state networks (templates obtained from \cite{Bassett2016}) and simulated RSNs using sICA. We use the Cohen's Kappa index $\kappa$ as a measure of similarity. 
(c) Probability distribution functions at the corresponding minimum of $\chi^2$, that is, $T/T_c=0.8$ and $T/T_c=1$ for the non normalized and normalized networks, respectively. 
The green line represents the empirical distribution. (d-f) Empirical and simulated functional connectivity matrices for the same parameters used in (c). The functional matrices 
are organized in blocks with RH (right hemisphere) and LH (left hemisphere).}
\label{FIG2}
\end{figure*}

We now investigate the consequences of these dynamical features on the simulated functional connectivity matrices. We employ the Pearson correlation and the $\chi^2$ 
distance to quantify the quality of our simulated averaged matrices (see Methods). The first index simply gives a linear correlation between the matrix elements, while 
the second one measures the distance between the two probability distribution functions. As usually done, we transform the model and empirical functional matrices (setting 
all diagonal elements to zero) in vectors, $F_m$ and $F_e$ respectively, and the Pearson correlation between both vectors, $\rho(F_m,F_e)$, is computed. The chi-squared 
distance is then calculated from the corresponding (normalized) probability  distribution functions $p(F_m)$ and $p(F_e)$ (see Methods). In Fig. \ref{FIG2} (a) we plot 
$\rho(F_m,F_e)$ as a function of $T/T_c$ for both $W$ and its normalized counterpart $\widetilde{W}$. The normalization of the excitatory input leads to drastic effects on the 
simulated functional matrices, thus suggesting the relevance of homeostatic principles in regulating brain functioning. Indeed, it enhances the correlation with the empirical data by a factor of $\sim 1.5$ at $T/T_c=0.6$ with respect to the non-normalized dynamics 
(see Figure \ref{FIG2}). We find that in both cases the best model performances $\chi^2 \approx 0.4$ and $\rho(F_m,F_e)\approx 0.6$ occur near the critical point $T=T_c$. How 
already stressed, due to finite size effects, for this database we have deviations from the thermodynamics ($N\rightarrow\infty$) critical point. It is interesting to compare the 
performance of our critical whole-brain model with a previous work by Deco {\it et. al.} \cite{Deco2013}, where a mean field approach has been employed to study the emergence of functional 
connectivities. Using the same structural input (averaged DSI matrix), but different functional data, we obtain $\rho=0.6$ (at $T/T_c=0.6$ using the balanced matrix $\widetilde{W}$) 
against $\rho=0.5$ of Deco {\it et. al.} best matching (see Fig. 3 in Ref. \cite{Deco2013}).

\begin{figure*}
\centering
\includegraphics[width=0.67\linewidth]{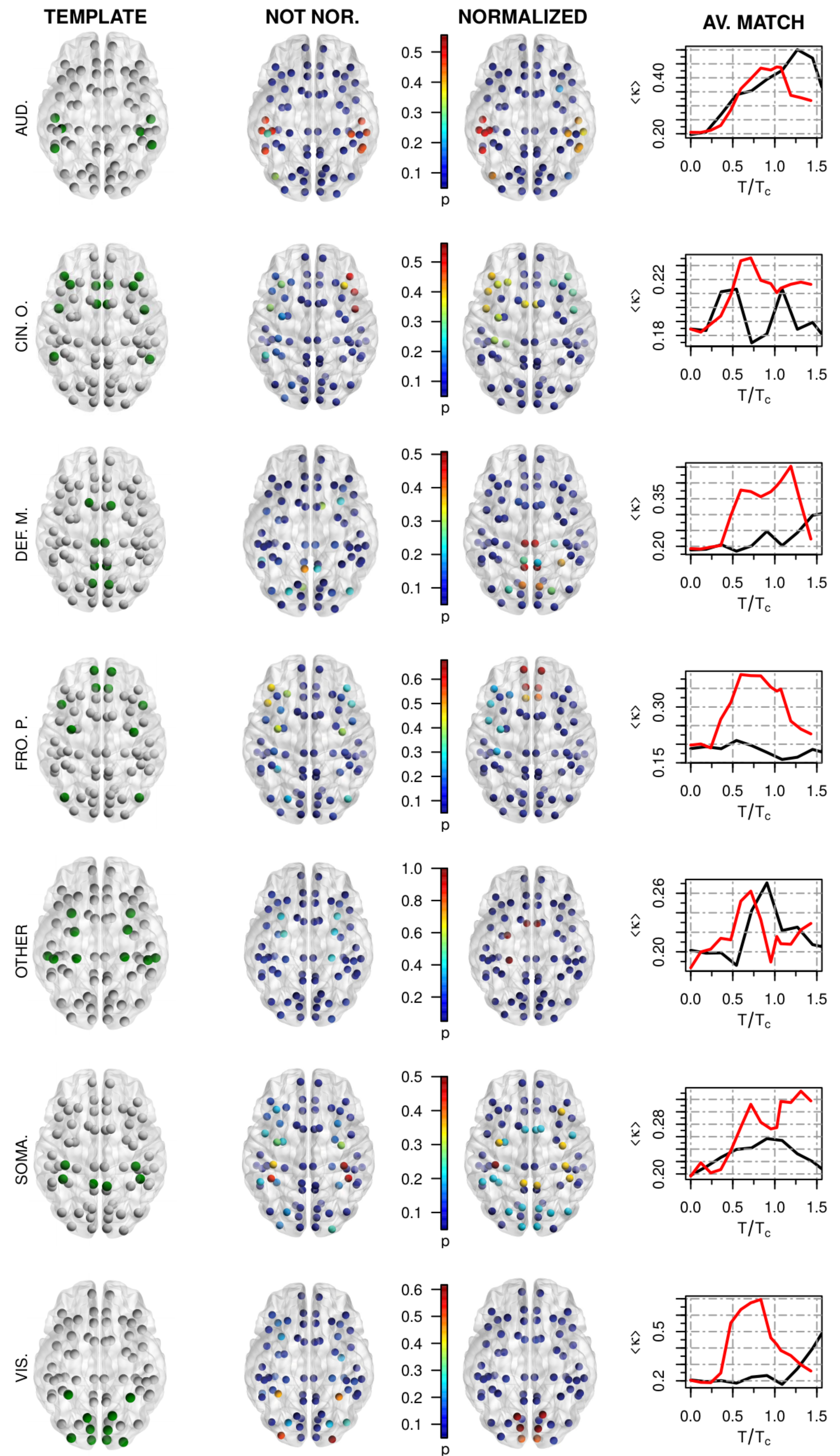}
\caption{In the first three columns we show a three dimensional projection of the RSNs considered during our analysis. First column: RSNs templates taken from \cite{Bassett2016}: 
auditory (6 matched nodes), cingulo-opercular (12), default-mode (8), fronto-parietal (10), “other” (10), somatosensory (6) and visual (10) networks. Second/Third columns: 
Simulated RSNs at the critical point $T_c$ given by the peak of $\langle S_2\rangle$ for the not-normalized and normalized structural networks respectively. Intensity values represent 
the probability, $p \in [0,1]$, that each node is included in the corresponding RSNs (spatial maps averaged over 100 trials). Therefore, nodes colored in blue do not belong to the 
corresponding RSN. Fourth column: Average match $\langle \kappa \rangle$ between simulated and template RSNs as a function of $T/T_c$ for the not-normalized (black) and normalized 
networks (red).}
\label{FIG3}
\end{figure*}

For the purpose of visual inspection, we show in Figs. \ref{FIG2} (d)-(f) the empirical connectivity matrix and the simulated ones at the corresponding minimum of $\chi^2$, 
that is, $T/T_c=0.8$ and $T/T_c=1$ for the non normalized and normalized networks, respectively. Notice that for these particular values of $T/T_c$ both distributions have 
approximately $\chi^2\approx 0.4$. The connectivity patterns predicted by the normalized model exhibits a balanced structure similar to what is observed in the empirical 
network (Fig. 2 (e)-(f)), further suggesting the role of homeostatic principles in capturing the topological features of the empirical network. Interestingly, such 
balanced connectivity structure is not present in the not-normalized FC.

To gain a deeper understanding of the effects of the (homeostatic) normalization we now analyze the brain organization into resting state networks (RSNs).
These are a set of areas in the resting brain, i.e., when the brain is not performing any specific cognitive, language, or motor tasks, displaying BOLD fluctuations that 
are correlated and synchronous within the same network \cite{Beckmann2006}. It has been found that RSNs are closely related to brain activation patterns seen during a given 
task execution, for instance, sensory (visual, auditory), cognitive, and motor etc. These spatiotemporal patterns can be obtained through spatial independent component analysis 
(sICA), that is the common statistical tool employed to extract RSNs from the BOLD activity (see Methods section). 

In Fig. \ref{FIG2} (b) we show the overall 
match between simulated RSNs using sICA and a template of well-established RSNs (taken from \cite{Bassett2016}) computed for the non-normalized ($W$) and the 
normalized ($\widetilde{W}$) networks as a function of $T/T_c$. We use the Cohen's Kappa index $\kappa$ as a measure of similarity (see Methods). We again observe that at 
$T_c$ the normalization performed on the structural connectivity matrix significantly increases the overall match with the empirical resting state networks. This result is 
consistent with the previous result shown in Fig. \ref{FIG2} (a). In addition, some simulated RSNs are more affected by the normalization, thus showing increased $\kappa$ 
similarity, as is the case of the default mode, the frontoparietal, the somatosensory and the visual networks (see the fourth column in Fig. \ref{FIG3}). 
In contrast, the auditory, the cingulo-opercular and the \textquotedblleft other\textquotedblright \, networks are not significantly affected by the normalization and both 
input structural matrices present similar $\kappa$ values. 
Further, we observe that some RSNs are more similar to the template in sub- or super-critical regions of the parameter space. Because of finite size effect 
(the notion of criticality is well defined only for infinite size systems), there is an intrinsic variability in the behavior of the model.
In Fig. \ref{FIG3} we also show a three dimensional projection of the templates and simulated RSNs at $T_c$. It is encouraging that a quite reasonable quality of the 
simulated resting state networks are achievable using a relatively low-resolution network. In summary, our findings support that inclusion of homeostatic 
principles successfully facilitates the formation of RSNs at criticality.

\textbf{Rudie et al. Dataset.} We extend now our analysis to other large-scale (open-access) structural and functional dataset of Rudie {\it et. al} \cite{Rudie_dataset}. 
Different from the previous case, the actual dataset contains a large number ($n=43$) of individual DTI/fMRI matrices, all of them parcellated in $N=264$ large-scale regions. 
By employing this dataset we can further investigate the issue of finite size effects stressed in the previous section and, most importantly, we can quantify the inter-subject 
variability presented by the neural patterns of brain activity. The only limitation is that we do not have a reference template to the RSNs, and therefore we have not 
considered them in our analysis. Here we present the results for the group level DTI/fMRI matrices (averaged over 43 healthy individuals). We let the analysis of personalized 
brain modelling to the next section. 

\begin{figure}
\centering
\includegraphics[width=\linewidth]{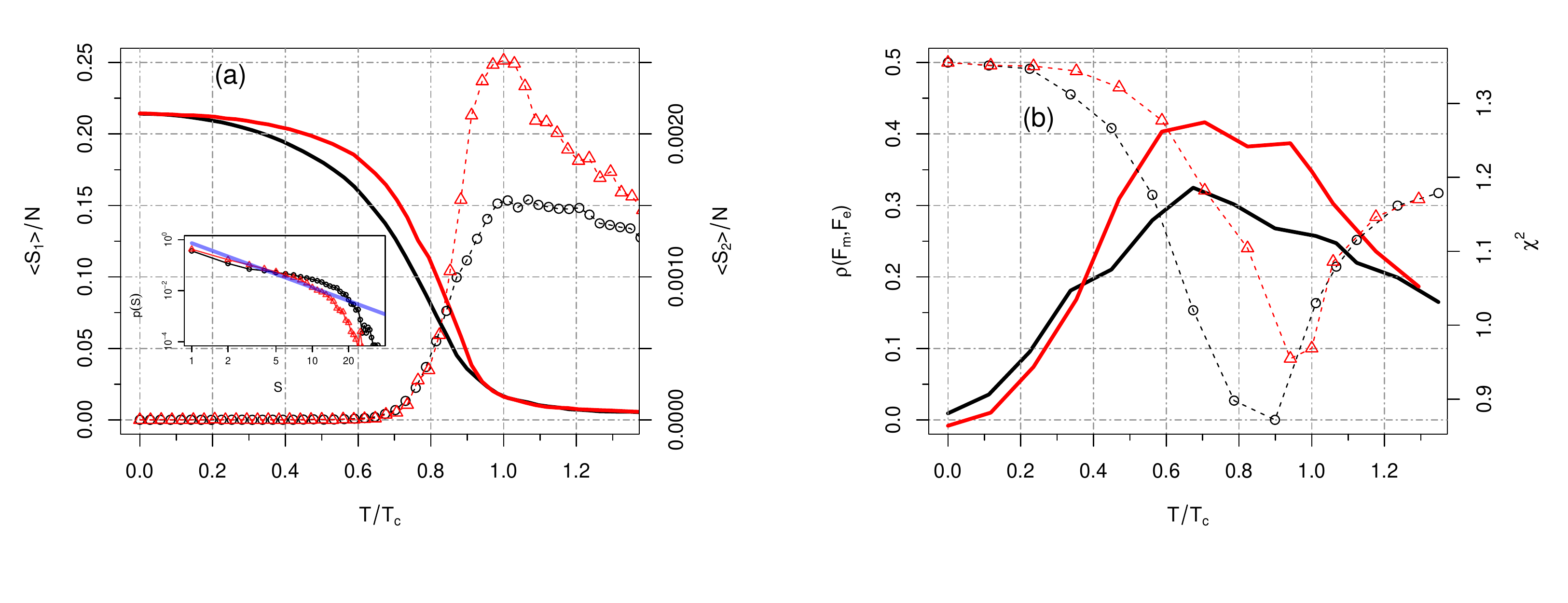}
\caption{Results of our framework applied to the Rudie {\it et. al} dataset \cite{Rudie_dataset}. (a) Largest (solid lines) and second largest cluster (dots) as a function 
of $T/T_c$. In black (red) we present the non normalized (normalized) system. Inset: Distribution of cluster sizes (in log-log scale) for the corresponding peaks of 
$\langle S_2\rangle$. The blue solid is the fit of Eq. \eqref{CDF} used to estimate $\alpha=1.74\pm 0.03$ for the normalized network. (b) Pearson correlation (solid lines) between 
simulated and empirical functional connectivity matrices, $\rho(F_m,F_e)$, as a function of $T/T_c$. Chi-squared distance (dots) between the corresponding (normalized) 
probability distribution functions. The critical points $T_c$, for all cases, were set as the peak in the second largest cluster because of the power-law distribution of 
cluster sizes found at this point.}
\label{FIG4}
\end{figure}

We fixed the two input parameters as in the previous dataset, i.e., $r_1=2/N$ and $r_2=r_1^{1/5}$. The total simulation time was set to $t_s=3,000$ time-steps, 
with time discretized in $dt=0.1$ seconds. Our results are shown in Fig. \ref{FIG4} (we maintain all the previous conventions about lines and colors as in Figs. \ref{FIG1}-\ref{FIG2}). Although 
the network size is $4$ times larger than the one in the previous dataset, we still find finite size effects. Nevertheless the critical point for $\langle S_2\rangle$ and 
$\sigma(A)$ are now closer. The non-normalized system presents a rather broad peak of $\langle S_2\rangle$ at $T_c$, while in the normalized case 
the peak is much sharper. Similar results hold also for the mean activity and its standard deviation with peaks located  at $T/T_c=0.8$ (not shown). 

As in the previous section, the distribution of cluster sizes for the normalized input matrix $\widetilde{W}$ displays a truncated power-law behavior in proximity and at the critical point 
with an exponent given by $\alpha=1.74\pm 0.03$ (see inset of Fig. \ref{FIG4} (a)). 

We finally investigate whether the equalization of the excitatory input increases the correlation between simulated and empirical functional connectivity matrices. Our results 
are shown in Fig. \ref{FIG4} (b). We find again that model performance is maximized near the critical point (due to finite size effect the maximum is not exactly at $T=T_c$) and 
the normalization of the network weights caused a substantial improvement of the simulated functional connectivity matrices. At the critical point, the Pearson correlation 
increases by a factor of $\sim 1.5$ with respect to the non-normalized input matrix. Regarding the chi-squared distance, eq.(\ref{ED}), both systems present almost the same performance at the 
critical point ($\chi^2 \approx 0.9$). 

\subsection*{Personalized brain modelling.}
By exploiting all the information in the Rudie {\it et. al} dataset, we can  quantify the variability of the critical points and neural activity patterns for different 
individuals and their dependence on the topological properties of the underlying individual connectomes. In order to address the above issues, we have simulated the stochastic 
dynamics for each individual in the dataset ($n=43$ healthy subjects) and calculated both the mean, the standard deviations and average clusters size of activity patterns. 
Simulations have been performed using the same model parameters of the group case.

\begin{figure*}
\centering
\includegraphics[width=\linewidth]{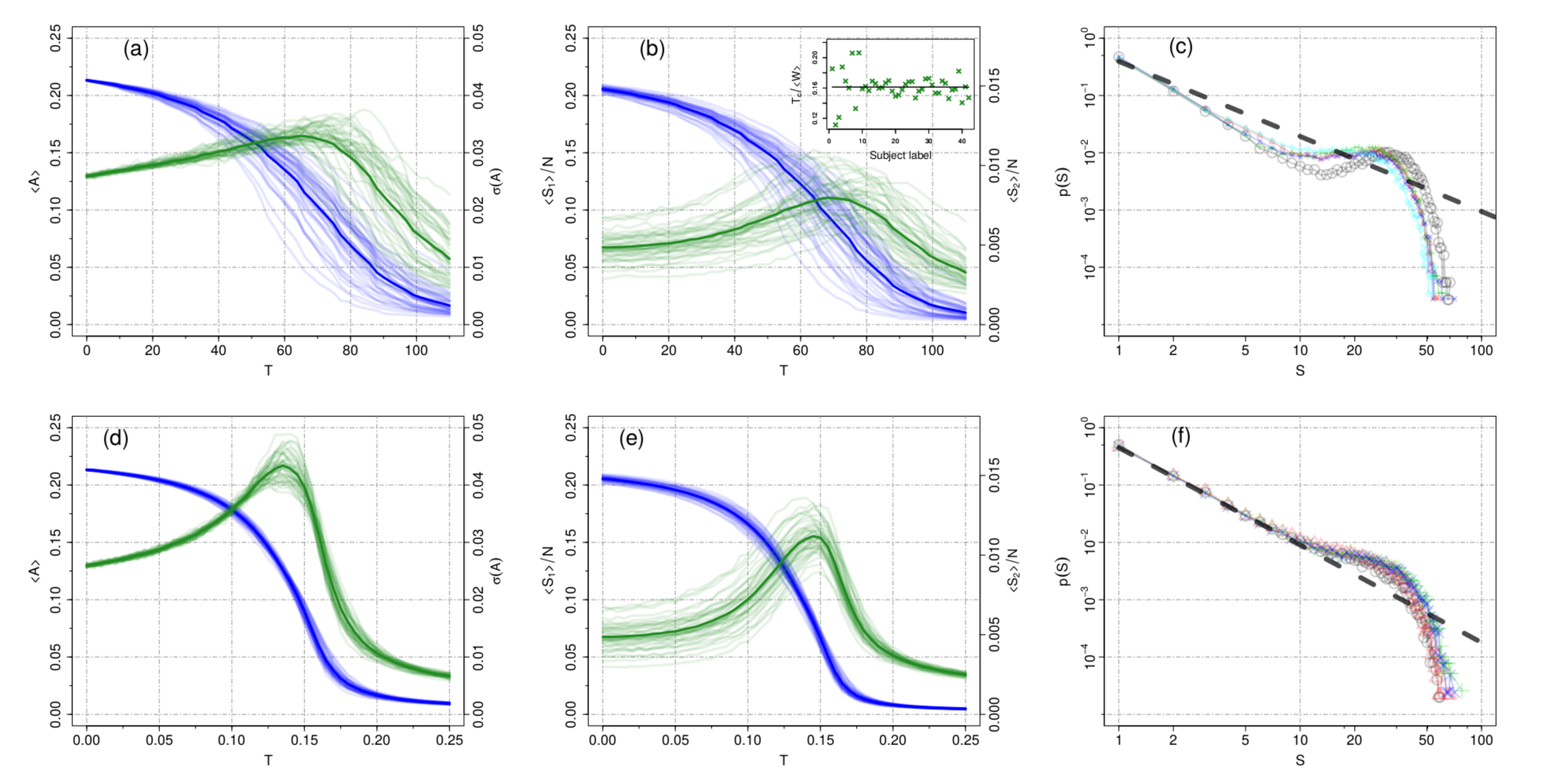}
\caption{Upper Panels: State variables as a function of the threshold for each participant of the Rudie {\it et. al} dataset for the non-normalized structural matrix, $W$. 
In (a) we show the mean activity (blue) and its standard deviation (green), while in (b) we show the largest (blue) and the second largest cluster (green). The thicker solid 
curves represent the group average. All individuals display a peak in both the standard deviation of the activity and in the second largest cluster vs $T$. The high 
variability in the location of the peak (critical points) is due to the individual average matrix entries $W_{ij}$, $\langle W \rangle$ (see main text). The collapse, at 
about $0.16$, of the ratio of the individual critical thresholds and of the individual average network strengths, $T_c/\langle W \rangle$, is shown in the inset of figure (b). 
(c) Cluster size distribution for some representative individuals at the corresponding critical point given by the peak of the second cluster size. Lower panels: the same as before but 
with the normalized matrix $\widetilde{W}$. At variance with the non-normalized system now we observe an almost perfect collapse of the individual's state variables, which 
in turn presents an enhanced phase transition all peaked close to $T_c\approx 0.15$. The shape of the distributions for the non-normalized systems is more distant from a 
power-law distribution than for the normalized ones. The exponents obtained by fitting Eq. \eqref{CDF} are:  $\alpha = 1.31 \pm 0.04$ and $\alpha=1.70 \pm 0.01$ for the non 
normalized and normalized systems, respectively. The value of the exponents corresponds to an average over 5 individuals.}
\label{FIG5}
\end{figure*}

We first analyzed the non-normalized networks. Fig. \ref{FIG5} (a)-(b) shows the behavior of $\langle A \rangle$, $\sigma(A)$, $\langle S_1\rangle$ and $\langle S_2\rangle$ 
vs $T$ for each of the participants. The heavy lines correspond to the average curve, e.g. $\langle A\rangle_{av}\equiv\sum_i^n \langle A\rangle_i/n$, where $n$ is the total 
number of participants. Consistent with the previous results, each of the four quantities displays a smooth behavior as a function of $T$. Furthermore, in the non-normalized 
case, each individual has its own critical threshold, according to the mean field prediction (see Methods). At the same time, consistent with the theory, the ratio between the 
critical threshold and the average strength for each individual connectome, does not change among individuals, i.e., $T_c^{(i)}/\langle W^{(i)}\rangle = k$, with $k$ a constant for $i = 1, \cdots,n$. 
In the inset of Fig. \ref{FIG5} (b) we show the ratio $T_c/\langle W\rangle$ for each of the $43$ participants. Except for few individuals, almost all points are peaked around 
$T_c/\langle W\rangle=0.161 \pm 0.017$, i.e. the dependence of $T_c$ on $\langle W\rangle$ is correctly captured by our mean-field approach.

Next we considered the normalized networks (see Figs. \ref{FIG5} (d)-(e)). The first striking result is the almost perfect collapse of $\langle S_1\rangle$, 
$\langle A\rangle$ and $\sigma(A)$ (the latter shows a small variability in the peak heights) of all analyzed subjects. Regarding the second largest cluster, although the 
individual curves do not perfectly collapse into each other, their peaks are sharply distributed around $T_c=0.15$. Thus importantly, as predicted by our mean-field approach 
(see Methods), for the normalized connectivity structure, the critical point of the dynamics is the same for all individuals in the dataset.

In order to test if the peak in the $\langle S_2\rangle$ curve for each individual is a marker of criticality, we have analyzed the distribution of cluster sizes at the corresponding critical 
points, using as input both matrices $W$ and $\widetilde{W}$. Figures \ref{FIG5} (c) and \ref{FIG5} (f) exemplify our results for some representative 
individuals $n=5$ (results do not change for other individuals). We obtain truncated power law distributions with a relatively small cutoff ($\approx 5$) with an averaged 
exponent $\alpha = 1.31$ for the non-normalized matrix, $W$, whereas for the normalized networks, $\widetilde{W}$, the power law extends about three times more (cutoff $\approx 15$) 
with an averaged exponent $\alpha=1.70$. The poor quality of the fit observed in Fig. \ref{FIG5} (c) suggest that scaling (if any) is less visible for the 
not-normalized network.

Finally, we have analyzed the performance, at the individual level, of the normalized HTC model with respect to the functional connectivity matrix (FC). Individual FC matrices are  
computed at the corresponding critical points, both for $W$ and $\widetilde{W}$, for some representative individuals ($n=10$). In all analyzed networks (results do not change 
for other individuals), the normalized HTC model shows a better performance. Indeed, we find an increased correlation between each simulated and empirical FCs by almost a 
factor of 1.5 as compared to the original HTC model. The mean of the two distributions is $\langle \rho_W\rangle = 0.111$ and $\langle \rho_{\widetilde{W}}\rangle= 0.161$ 
($p$-value$=0.0004$) for the non-normalized and normalized networks respectively. Together these results demonstrate that the inclusion of homeostatic 
principles generate more realistic model prediction on brain dynamics akin to criticality, suggesting that the normalized model is indeed better than its non-normalized 
counterpart.

\section*{Discussion}

In this work, we have explored how the inclusion of homeostatic plasticity principles affects the macroscopic dynamics and the formation of functional networks of a previously studied stochastic whole-brain model \cite{Hai2013}. 
In the simulated dynamics, the long-distance connections linking the mesoscopic brain regions have been described by the structural network defined by the human connectome. 
Homeostatic plasticity has been implemented in a static way as a normalization of the incoming node’s excitatory input. %, and it has the main effect to balance the macroscopic dynamics. 

Our main findings are:

\begin{enumerate}

\item Normalization of the node's excitatory input leads to a robust neuronal dynamics consistent with the hallmarks of criticality. It successfully balances the macroscopic 
dynamics, and it increased the strength of spatiotemporal fluctuations. The clusters of activity, $S(t)$, became more heterogeneous spreading out along the whole 
network and not along the hubs as in the not-normalized model. In other words, normalization increased the strength of the critical transition (i.e., increased $\sigma(A)$ 
and $\langle S_2 \rangle$), suggesting that a better representation of the macroscopic brain dynamics has been reached.

\item In the normalized model, the cluster size distribution, in proximity to the critical point, followed a truncated power-law with a critical exponent $\alpha$ close to the 
hallmark exponent of avalanches sizes, $\alpha = 3/2$. On the other hand, scaling invariance in the cluster size distribution (if any) is less visible for the 
not-normalized model. The emergence of a critical-like dynamics due to the inclusion of homeostatic principles has also been investigated in \cite{Leech2015}. 
The implemented inhibitory homeostatic plasticity has brought the macroscopic dynamics close to criticality, and avalanches sizes following power-law distribution has 
been observed. Our results further suggest the fundamental role of homeostatic processes to bring the dynamics of the brain close to a critical state.

\item We have found, in accordance with similar studies \cite{Corbetta2014,Leech2015,Woolrich2018}, that inclusion of homeostatic principles significantly improves the 
correspondence between simulated and empirical functional networks based on fMRI. At the critical point, the functional connectivity patterns predicted by the normalized 
model exhibited a more balanced structure similar to what is seen in the empirical network (compare Fig. 2 (e)-(f)), suggesting that the homeostatic model better predicted 
the topological features of the empirical functional networks. Interestingly, such a balanced structure was not present in the not-normalized FC. Furthermore, the homeostatic
model presented a significant increase in correlation coefficients, by a factor of $\sim 1.5$, as compared to the not-normalized model. We have also considered the size effect 
of the brain networks and we have not found any significant difference in the correlation coefficients, at least for the network sizes considered in the present study. This 
result goes in contrast with the ones in reference \cite{Messe2015} were significant size effects on the predictive power has been reported.

\item Simulated resting state networks exhibited more realistic spatial patterns in presence of homeostatic plasticity principles. We have obtained RSNs maps through sICA and we have 
compared them with a template of RSNs \cite{Bassett2016}. Our results suggest that the inclusion of homeostatic principles successfully facilitates the formation of RSNs at 
criticality. Therefore, we have observed a significant increase in the correspondence between simulated RSNs and the template, as attested by the Cohen's Kappa similarity 
index. However, we have found that some resting state networks were more similar to the template in sub- or super-critical regions of the parameter space. Because the 
relatively low-resolution network used to extract the RSNs, there is an intrinsic variability in the behavior of the model.

\item Normalization minimizes both the variability of the critical points and neuronal activity patterns among healthy subjects. In particular, we have shown that normalization collapses 
the model state variables, i.e. neural activity patterns, of healthy subjects into universal curves. We have demonstrated these results employing a 
combination of analytical and numerical tools. Indeed, we have written a mean-field version of the macroscopic dynamics. We have shown that our mean-field solution accounts, with a reasonable agreement, for the variability of the 
critical points with the network strength $\langle W\rangle$ observed numerically. Indeed, as predicted by our mean field solution, for the normalized dynamics the critical 
point is the same for all individuals. Finally, from a theoretical point of view, the in-degree normalization is a necessary step to ensure that 
the critical threshold ($T_c$) does not depend on the size of the system.

\end{enumerate}

In this study, we have hypothesized that at the whole-brain level the brain network would already be balanced thanks to homeostatic plasticity mechanisms regulating the 
interplay between excitation and inhibition, and we have explored how this feature affected the macroscopic dynamics and the formation of functional networks. Overall, we 
have observed a significant increase in correlation coefficients, resulting in more realistic model predictions. There are strong experimental and theoretical evidence 
supporting that homeostatic plasticity mechanism, across spatiotemporal scales, are crucial for regulating neuronal and circuit excitability \cite{Desai2010,Turrigiano2011,Xu2015}. 
In particular, recent theoretical works suggest that inhibitory synaptic plasticity (ISP) may provide a plausible homeostatic mechanism to stabilize neuronal dynamics at 
the whole-brain level. Such balancing between excitation and inhibition has been demonstrated through a biophysical Wilson-Cowan modeling framework on fMRI 
\cite{Leech2015,Corbetta2014} as well as MEG timescales \cite{Woolrich2018}. Despite the fact that the HTC model does not consider local plasticity at the 
inhibitory-excitatory connections like in \cite{Leech2015,Corbetta2014,Woolrich2018}, normalization of the in-degree has the basic effect to adjust locally the network 
inhibition and that it renders activity levels approximately constant across the brain regions. Indeed, the normalization rule adopted in this work resembles, at the 
macroscopic level, the synaptic scaling \cite{Nelson1998,Royer2003}, which describes the up (down) 
regulation of a neuron's synaptic input in order to keep its firing rate within some target range.

One of the important problem that theoretical neuroscience needs to tackle is personalized brain modeling \cite{Ana2016,Bartolomei2017,Bansal2018}. 
A key role in this challenge is played by whole brain models, which are grounded on the study of the human brain as a dynamical, complex and self-organized 
networked structure. Indeed, ideally, the brain activity derived from whole brain models should predict functional recovery in patients who have suffered brain damage 
(e.g. due to stroke). However, this attempt is strongly limited by the fact that the strength of the (cor)relations between model activity and data at the level of the 
individual subject \cite{Deco2014} is very modest and the predictions can be very inaccurate. For this reason, most often whole brain models have their parameters tuned 
so at to replicate certain functional indicators at the population level, e.g. functional connectivity computed using the correlation between observed time series \cite{Friston1996} 
or spatiotemporal patterns of local synchronization \cite{Deco2011a}. In particular, most of the models have developed indexes that are able to distinguish between different 
groups of subjects (e.g. healthy vs. stroked) using as input an average connectome obtained from many individuals and comparing the model output with average group properties \cite{Hai2016,Tononi2015}. 
However, a prerequisite for theoretical models to be significant in terms of translational neuroscience, and thus possibly informative for therapeutic intervention, is to 
provide individual-level markers (or \textquotedblleft brain signatures\textquotedblright) that reliably predict cognitive and behavioral performance not only at the group 
level but also be adapted and tailored to the specific patient. Different markers have been recently proposed, for instance, information capacity \cite{Adhikari2017} (an 
information theoretical measure that cannot be obtained empirically), integration \cite{Adhikari2017} (a graph theoretical measure obtained from functional connectivity) and 
entropy \cite{Saenger2017} (a measure of repertoire diversity). All such quantities showed decreased values in stroke patients \cite{Adhikari2017,Saenger2017}, while 
information capacity and integration were additionally correlated with measures of behavioral impairment \cite{Adhikari2017}. The relation between these measures and 
criticality will be tackled in a future work.

In the present study, we have shown that by introducing homeostatic principles to the HTC model, we are able not only to generate realistic brain dynamics at the group level but 
also to provide individual-based markers that reliably predict neuronal state activity, i.e. the criticality of the brain. In each of the analyzed connectomes of the Rudies 
et al. dataset, we, in fact, have found that the critical point of the generalized HTC model is located at the same $T_c$. Moreover, both $\langle S_2\rangle$ and $\sigma(A)$ 
have a maximum at this parameter value, and the system displays long-range correlations. A recent paper by Haimovici {\it et. al.} \cite{Hai2016} investigated the effect of 
artificial lesions on the signatures of criticality of the network dynamics. Lesions were simulated  by removing nodes/links of an averaged group level structural empirical 
matrix, targeting nodes/links according to a given network centrality and also in a random way. They found that stroked simulated brains have shifted values 
of the critical point with respect to the  healthy reference point, i.e., synthetic lesions brought the system to a sub-critical state, which is characterized by decreased levels of
neural fluctuations (i.e, decreased mean activity $\langle A \rangle$ and standard deviation $\sigma(A)$). Sub-critical dynamics also lead to alterations in the functional 
parameters, for instance, the mean and variance of the functional connectivity matrix are decreased due to synthetic lesions. Other studies have reported decrease in 
long-range correlations in neural activity during anesthesia \cite{Scott2014}, slow wave sleep \cite{Priesemann2013} and epilepsy \cite{Meisel2012}.
However, such an approach would not be applicable in models where the critical point is individual dependent. Our results for the normalized HTC 
model show that introducing an equalization of the excitatory input in the simulated brain dynamics minimize the variability of the neural activity patterns and the critical point 
of the HTC model for different (healthy) subjects, allowing the opportunity of statistical comparison among model outputs for single individuals. 
We believe the inclusion of homeostatic principles in the HTC model not only will reproduce known emerging patterns in stroke (or other brain disorders) 
but could eventually discover new ones.

Despite the methodological novelty of the presented model is limited, simply consisting in adding a homeostatic normalization to the HTC model, 
the manifested effects clearly lead to better representations of the macroscopic brain activity and it is an improvement over the previous model. As future perspectives, we 
wish to investigate the influence of other parameters in the model, for instance, the effect of a continuous transfer gain function, while a recent work \cite{Poldrack2018} has shown that 
modulation of the response gain parameter (the Heaviside function in Eq. \eqref{p}) can mediate a critical transition in the brain. Also, similar to what was 
done in \cite{Leech2015}, we wish to implement homeostatic plasticity by adding a local dynamics on the node's threshold $T_i(t)$.

In summary, network normalization is useful in increasing the spatiotemporal variability of the brain dynamics at the individual level, which in turn increases the correlation between models 
outputs and empirical data. When applied at individual connectomes, the model collapses the state variables of healthy subjects into universal curves. A natural follow up of 
this work will be to develop an individual-level marker based on criticality (calculated as $\langle S_2(T_c) \rangle$ for example) that reliably predict cognitive and 
behavioral performance (like in \cite{Adhikari2017,Saenger2017}) as well as its evolution following therapeutic intervention. 
In particular, the main application we have in mind is the study of brains affected by stroke. For instance, we are interested in investigating how anatomical damage 
could affect brain’s critical dynamical regime and underlying functional organization. The modeling of real stroked connectomes in light of criticality remains almost 
unexplored. The reduced inter-subject variability of the normalized HTC model is its key feature. Indeed, it allows the opportunity of statistical comparison among model 
outputs for single individuals, then opening new perspectives to study stroke recovery using empirical DTI/fMRI data of single stroke patients.

\section*{Methods}

\section*{Empirical datasets of structural connectivity and functional networks}

We analyzed two different datasets of healthy subjects, consisting on both functional (fMRI) and structural (DTI/DSI) data. Specific details on the data acquisition and 
preprocessing can be found in the original studies.

The Hagmann {\it et. al} dataset consists of a group level (DSI) structural matrix averaged over five healthy subjects \cite{Hag2008}, 
and parcellated in $N=66$ cortical regions. The entries of the connectivity matrix $W$ represents the number of connecting fibers between a given pair of regions 
divided by the average area and by the average fiber length between the two regions. Functional data corresponds to BOLD time-series measured from a cohort of other 
$24$ healthy subjects taken from Corbetta et. al \cite{Corbetta_dataset}. Each subject performed two scanning runs of 10 minutes at rest. We used the Pearson correlation, 
eq. \eqref{Pearson}, to compute the FC matrix of each subject/scan. Then we averaged all FC matrices to obtain the group level FC. 

The Rudie {\it et. al} dataset \cite{Rudie_dataset} consists of structural (DTI) and resting state functional matrices (obtained with BOLD fMRI) parcellated in $N=264$ cortical 
regions from a cohort of $n=43$ healthy typically developing individuals (13.1 $\pm$ 2.4 years). In this case, the entries of the connectivity matrix $W$ represents the total 
number of fibers connecting a given pair of regions. To obtain the group level structural (functional) matrix we computed the average over the entire group of $n=43$ participants. 

During our numerical experiments, we have analyzed the two datasets in different ways. In the first two parts of our analysis, we have investigated the effects of network 
normalization on the neural patterns at the group level, and thus we have simulated dynamics with an averaged structural network. Then we have compared the model output with 
the corresponding empirical group level FC. Finally, in the third part, we have investigated the inter-subject variability simulating the neural dynamics on each individual 
structural matrix of the Rudie {\it et. al} dataset, then showing the feasibility of personalized modelling using our approach.

\subsection*{Characterization of simulated brain activity}
In order to characterize the simulated brain activity through the generalized HTC model as a function of the control parameter $T$, we have considered the following standard 
quantities (for simplicity we will refer to them as state variables):
\begin{itemize}
 \item  the mean network activity,
\begin{equation}
\langle A \rangle = \frac{1}{t_s}\sum_{t=1}^{t_s} A(t),
\end{equation}
\item the standard deviation of $A(t)$,
\begin{equation}
\sigma (A)= \sqrt{\frac{1}{t_s}\sum_{t=1}^{t_s}\big( A(t)-\langle A \rangle\big)^2},
\end{equation}
where $A(t)=\sum_{i=1}^{N}s_i(t)/N$, $N$ is the total number of nodes and $t_s$ is the simulated total time; 
\item the sizes of the averaged clusters, the largest $\langle S_1\rangle$ and the second largest $\langle S_2 \rangle$. Clusters were defined as ensembles of nodes that 
are structurally connected to each other and simultaneously active. 
\end{itemize}

During simulations we kept fixed the model parameters of $r_1$, $r_2$ and $T$. Then we updated the network states, starting from random configurations of $I$ and $R$ states, 
for a total of $t_s$ time-steps. For each value of the threshold $T$ we computed the state variables, $\langle S_1\rangle$, $\langle S_2\rangle$, $\langle A\rangle$ 
and $\sigma(A)$. Throughout this study, unless stated otherwise, the final numerical results presented were averages over $100$ initial random configurations. 

\subsection*{Mean-Field prediction of the critical point}
Analytical solutions for $p_i^t$ and $q_i^t$ of Eqs. \eqref{p}-\eqref{q} are difficult to be obtained. However, by studying the stead-state solution 
($t\rightarrow\infty$) of the mean-field approximation we are able to explain the inter-subject variability of the critical points. Indeed in the stationary state by setting 
$p_i^t=\bar{p}_i =\bar{p}$ and $q_i^t = \bar{q}_i =\bar{q}$ in Eqs. \eqref{p}-\eqref{q} and approximating $W_i\equiv \sum_j W_{ij}\approx\langle W\rangle$, where 
$\langle W \rangle =\sum_i W_i/N$ is the average network strength, one obtains, after straightforward manipulations,
\begin{equation}
T_c =\langle W \rangle  r_2 /(1 + 2r_2),
\label{MFS}
\end{equation}
as the critical point. One finds that $\bar{q}=\bar{p}= r_2/(1+2r_2)\equiv p_{-}$ when  $T<T_c$ and $\bar{p}=r_1\bar{q}= r_1r_2/(r_1+r_2+r_1r_2)\equiv p_{+}$ when $T>T_c$. 
Notice that $p_{+} < p_{-}$ as one would expect, i.e. the activity is large when the threshold is low. This expression is only an approximation of the exact critical 
threshold. However, the mean-field solution accounts, with a reasonable agreement, for the variability of the critical points with the network strength $\langle W \rangle$ 
observed numerically. Importantly, we see that for the non-normalized dynamics, $T_c$ depends on the specific individuals, as $\langle W \rangle$ is different among 
different brains. On the other hand, when using normalized structural connectivity then $\langle \widetilde{W} \rangle = 1$ for all individuals and therefore $T_c$ is universal.

\subsection*{Model Validation}
\textbf{From the Model Output to BOLD Signal.} Experimentally, brain activity at rest can be accessed through fMRI. In fMRI what is measured is the variation of the 
blood-oxygen-level dependent  (BOLD) signal. Moreover, following \cite{Hai2013} we simulate BOLD time-series of each node convolving the node variable 
$s_i(t)$ with a canonical double-gamma hemodynamic response 
function (HRF), 
\begin{equation}
x_i(t)=\int_{0}^{\infty}s_i(t-\tau)h(\tau)d\tau,
\end{equation}
with,
\begin{equation}
\displaystyle h(\tau)=\Big(\frac{\tau}{d_1}\Big)^{a_1} e^{-\frac{\tau-d_1}{b_1}} - c\Big(\frac{\tau}{d_2}\Big)^{a_2} e^{-\frac{\tau-d_2}{b_2}}, 
\label{functionH}
\end{equation}
where $x_i(t)$ is the BOLD signal of the $i$-th node. The free parameters in \eqref{functionH} were fixed according to values found in \cite{Glover99}, i.e., $d_i = a_i b_i$, 
$a_1 = 6$, $a_2 = 12$, $b_i = 0.9$, and $c = 0.35$. Finally, the convolved time-series, ${\bf x}(t)$, were filtered with a zero lag finite impulse response band pass filter 
in the frequency range of $0.01-0.1$ $Hz$. Although complicated, these steps are part of a standard procedure to transform model output in BOLD functional signals.

From the generated BOLD signal we can finally  extract the functional connectivity (FC) networks. In fact, the FC matrix $r_{ij}$ is defined through Pearson correlation:
\begin{equation}
r_{ij}=\frac{\langle x_i x_j \rangle- \langle x_i \rangle \langle x_j \rangle}{\sigma_i \sigma_j},
\label{Pearson}
\end{equation}
where $\sigma_i=\sqrt{\langle x_i^2 \rangle - \langle x_i \rangle^2}$ is the standard deviation and $\langle \cdot \rangle$ is the temporal average of the BOLD time series. 

To access the quality of our results we need to compare the generated FC matrix with the functional networks obtained from the fMRI data. In particular, we employed two 
distinct statistical measures to quantify the similarity between simulated and empirical functional matrices: (i) the Pearson correlation and, (ii) the chi-squared 
distance ($\chi^2$). As usually done, we transform the model and empirical functional matrices (setting all diagonal elements to zero) in vectors, $F_m$ and $F_e$ respectively, 
and the Pearson correlation between both vectors, $\rho(F_m,F_e)$, is computed. The $\chi^2$ distance is then calculated from the (normalized) probability 
distribution functions $p(F_m)$ and $p(F_e)$,
\begin{equation}\label{ED}
\chi^2 = \sqrt{\sum_i^{N_b} (p_i(F_m) - p_i(F_e))^2 / (p_i(F_m) + p_i(F_e))},  
\end{equation}
where $N_b$ is the number of bins used to calculate both histograms. 

\textbf{Resting State Networks.} The rest brain activity displays coherent spatiotemporal activation patterns which have been consistently found in healthy subjects 
\cite{Raichle2005,Beckmann2006}. These spatiotemporal maps reflect regions that are functionally connected, i.e., with a similar BOLD activity, although they 
may be anatomically disconnected. The brain organization into resting state networks have been vastly extracted using spatial and temporal independent component analysis 
(sICA/tICA) \cite{Ica1,Ica2}. Here we applied the spatial ICA (sICA) to extract RSNs from the BOLD activity. sICA decomposes a set of BOLD time-series into a number of $n$ 
independent components (specified a priori) which are spatial maps associated with the time courses of the signal sources. In matrix notation it reads,
\begin{equation}
Y = A S,
\end{equation}
where $Y$ is the ($ N_t \times N$) raw matrix containing in its columns the simulated time-series (of length $N_t=dt \cdot t_s$). Spatial maps are encoded in the rows 
of $S$ (of order $n\times N$) and the corresponding time courses of the signal sources in $A$ (of order $N_t \times n$).

We employed the fastICA algorithm in R (open-source platform) to estimate the independent components (ICs). After that ICs maps were z-transformed, i.e., 
$S_i^\prime = (S_i-\langle S_i\rangle)/\sigma(S_i)$ for $i=1,\cdots,n$. For each value of $T$ we repeated such procedure $100$ times with distinct initial random 
configurations. At the end, we end up with a pool of $100\cdot n$ ICs maps following a Gaussian distribution with zero mean and unity standard deviation. We finally 
threshold and binarize ICs. In particular, we set to 1 all elements such that $|S_i^\prime| \ge \theta$ ($i=1,\cdots,100\cdot n$) and zero otherwise. 

We access the quality of our simulated spatial maps (ICs) by computing the Cohen's kappa similarity index \cite{CohenKappa}, $\kappa$, with a template of well-established 
human RSN taken from \cite{Bassett2016}. The template contains the name of the anatomical brain regions mainly involved in a given RSN network. We used it to match each node 
of our $N=66$ network belongs to a given empirical RSN. Such procedure resulted in a total of $7$ binary RSN template vectors, namely, auditory (6 matched nodes), 
cingulo-opercular (12 nodes), default mode (8 nodes), fronto-parietal (10 nodes), somatosensory (6 nodes), visual (10 nodes) and \textquotedblleft other\textquotedblright\, 
(10 nodes) resting state networks. We omitted the ventral and dorsal attention templates because they comprised a small number of matched nodes ($2$). 
Following \cite{Hai2013,Glomb2017}, we performed a best match approach to assign each simulated IC to be belonging to a given RSNs. Indeed, we computed $\kappa$ between 
each $S_i^\prime$ with all template vectors, assigning the RSN with highest $\kappa$ and averaging the corresponding values across the pool of ICs to obtain the 
overall average match $\langle \kappa \rangle$ (see Fig. \ref{FIG2} (b)). We also computed $\langle \kappa \rangle$ for each RSNs by simply averaging those ICs assigned 
to be closest to a given template vector (see Fig. \ref{FIG3}).

We finally fixed the free parameters, namely, the threshold $\theta$ and the number of independent components $n$, in a data driven-way. Accordingly, we applied the above framework 
to the Corbetta et. al. dataset \cite{Corbetta_dataset} (48 empirical BOLD time-series) and then fixed the parameters in such a way to maximize the overall match $\langle \kappa\rangle$. By 
varying a two-dimensional parameter space we found a maximum at $n=8$ and $\theta$ corresponding to the 92-th percentile of the entire pool of $8\cdot 48$ ICs (see Supporting Figures, Figs. 
S1, S2 and S3).   

{\textbf{Fitting procedure.} Following \cite{Schappo2016} we use the complementary cumulative distribution function, $F(S)$, to perform our fits. We 
assume a power-law distribution $P(S)\sim S^{-\alpha}$ $(S \le Z)$ with a cutoff $Z$, therefore,
\begin{equation}
F(S) \equiv \int_S^Z P(S^\prime)d S^\prime = c_1 +c_2S^{1-\alpha},
\label{CDF}
\end{equation}
where the parameters $c_1$, $c_2$ and $\alpha$ are fitted to the complementary cumulative distribution using all the data points. The power-law exponents are computed from 
an average over 10 fits using different initial random configurations, each one lasting $t_s=15,000$ time steps. The cumulative distribution provides a clearer way to calculate the power-law 
exponent $\alpha$ because $F(S)$ can be directly obtained from the data, and it does not suffer from (binning) histogram estimates like $P(S)$. For the fitting procedure 
we use standard nonlinear least squares algorithm provided by R. As is customary in the field, for presentation purposes, we showed $P(S)$ in log-log scale (instead of $F(S)$).

\section*{Acknowledgements}
R.P.R. was supported by National Council for Scientific and Technological Development (CNPq Grant No.201241/2015-3) and the Research, Innovation and Dissemination 
Center for Neuromathematics (FAPESP Grant No. 2018/08609-8). R.P.R. thanks Francesco D'Angelo for useful discussions. L.K. Acknowledges the financial support of the 
Cariparo Foundation. M.C. was supported by NIH RO1NS095741. A.M. was supported by excellence project 2017 of the Cariparo Foundation.

\section*{Author contributions statement}

R.P.R, S.S., M.C. and A.M. designed the research, R.P.R, L.K. and S.S. performed the research. All authors wrote and reviewed the article.

\section*{Additional information}

{\bf Competing Interests:} The authors declare that they have no competing interests.


\begin{thebibliography}{99}
\bibitem{obs1} Strictly speaking phase transitions exist only for systems with an infinite number of degrees of freedom, which at best are good approximation of large, 
but finite, systems like a brain.
\bibitem{obs3} Based on the HTC model, the excitation-inhibition mechanism is solely dependent on the weighted in-degree; there is no physical meaning in 
normalizing the out-degree. If we consider a non-weighted network (i.e, a binary adjacency matrix), we still would observe the same emergent patterns from the model, at the condition that we normalize the in-degree.
\bibitem{obs2} Notice that a peak in a finite system does not necessarily is indicative of a transition in the corresponding infinite system. However in our case we make the assumption that the observed 
peaks in both $\sigma(A)$ and $\langle S_2 \rangle$, in the normalized dynamics, correspond indeed to a critical phase transition in the infinite system.
\end{thebibliography}
\end{document}